\documentclass[pra,aps,showpacs,superscriptaddress,twocolumn]{revtex4-1}
\usepackage{amsmath}
\usepackage{amsfonts}
\usepackage{amssymb}
\usepackage{mathrsfs}
\usepackage{epsfig}
\usepackage{color}
\usepackage{epstopdf}

\begin{document}

%------------------------------------------------Alternative commands-------------------------------------------------
\newcommand{\ket}[1]{\vert#1\rangle}
\newcommand{\bra}[1]{\langle#1\vert}
\newcommand{\braket}[2]{\langle#1\vert#2\rangle}
\newcommand{\ketbra}[2]{\vert#1\rangle\langle#2\vert}
\newcommand{\braketM}[3]{\langle#1\vert#2\vert#3\rangle}
\newcommand{\modq}[1]{\vert#1\vert^2}
\newcommand{\modulo}[1]{\vert#1\vert}
\newcommand{\mean}[2]{\langle#1\rangle_{#2}}
\newcommand{\vet}[1]{\overrightarrow{#1}}
%-----------------------------------------------------------------------------------------------------------------------------

\title{Vortex entanglement in Bose-Einstein condensates coupled to Laguerre-Gauss beams }

\author{N. Lo Gullo}
\affiliation{Department of Physics, University College Cork, Cork, Republic of Ireland}
\author{S. McEndoo}
\affiliation{Department of Physics, University College Cork, Cork, Republic of Ireland}
\author{T. Busch }
\affiliation{Department of Physics, University College Cork, Cork, Republic of Ireland}
\author{M. Paternostro}
\affiliation{School of Mathematics and Physics, Queen's University, Belfast BT7 1NN, United Kingdom}

\date{\today}

\begin{abstract}
We study the establishment of vortex entanglement in remote Bose Einstein condensates. We consider a two-mode photonic resource entangled in its orbital angular momentum (OAM) degree of freedom and, by exploiting the process of light-to-BEC OAM transfer,  demonstrate that such entanglement can be efficiently passed to the matter-like systems. Our proposal thus represents a building block for novel dissipation-{free} and long-memory communication channels based on OAM. We discuss issues of practical realizability, stressing the feasibility of our scheme and present an operative technique for the indirect inference of the set vortex entanglement. 
\end{abstract}
%\bigskip
%\keywords{BEC, vortices, entanglement transfer, orbital angular momentum, qudits}
\maketitle

\section{Introduction}

The rich variety of coherently exploitable degrees of freedom with which a photonic 
system is endowed has been extensively used in recent years in order to 
demonstrate the building blocks of quantum technology protocols including quantum 
cryptography~\cite{crypto}, quantum repeaters~\cite{repeaters}, teleportation and quantum computing~\cite{computing}. In this context, the exploitation of orbital angular momentum (OAM) carried by light is settling as a new and exciting opportunity for coherent manipulation at the classical and quantum level~\cite{review}. High density data transmission~\cite{high}, activation of micro-machines and optical tweezers~\cite{tweet} are among the most prominent applications of optical OAM so far. 

In addition, the field of quantum information processing has now started exploiting the additional opportunities offered by this photonic degree of freedom for communication and manipulation purposes. It has been shown that it is possible to create OAM-entangled photons by means of a {\it routinely used} setup such as spontaneous parametric down
conversion (SPDC)~\cite{entphot}. This has triggered a plethora of studies on how to generate, manipulate and detect non-classical states of OAM~\cite{torner}, culminating in the demonstration of Bell's inequality violation by OAM-entangled two-photon states~\cite{entphot2}, the introduction of so-called hyper-entangled states~\cite{hyper}, the design of quantum cryptographic schemes based on higher-dimensional systems~\cite{simon} and the transfer of OAM states from light to matter-like systems~\cite{atomi}.

In particular, the latter scenario holds the potential for the realization of experimentally feasible long-time quantum memories embodied by superfluid rotational states of Bose-Einstein condensates (BECs)~\cite{atomi,marzlin,than,dow,meystre}.  
%On the other hand, photons alone are not enough, we also need to store
%information in matter devices in order to process it or just to keep
%it until it is needed. Many systems are under study both theoretically
%and experimentally to achieve this goal. One very promising such
%system is the Bose-Einstein condensate (BEC) due to their macroscopic
%quantum nature. BECs offere a large number of bosons with high spatial
%and time coherence, for this reason they have been proposed as atomic
%lasers sources. 
The spatial coherence intrinsic in a BEC allows for a superfluid vortex state 
in which the bosons in the condensate have a well defined and quantized OAM, which 
offers a perfect match with rotating photon carriers. Along the seminal lines traced by the experiments in Refs.~\cite{atomi}, a few theoretical proposals for the light-to-vortex state transfer have been presented~\cite{marzlin,dow,than,meystre}. Here we close the circle of these proposals and show that it is possible to create entanglement between two spatially separated BECs by transferring OAM from entangled photon resources to the condensates. We propose a simple and efficient scheme to achieve this goal using experimentally achievable parameters and routinely produced OAM-entangled light resources.
%, together with a method for the achievement of maximal entanglement shared by two remote vortices. 
On a different level, our study proposes a scheme that is able to transfer (with in principle 100\% efficiency) higher-dimensional entanglement between two independent system by means of bilocal interactions, thus contributing to an area that is witnessing theoretical and experimental interest (see Choi {\it et al.} in~\cite{repeaters} and Ref.~\cite{mio}).

The paper is organized as follows. In Sec.~\ref{modello} we introduce the idea behind our proposal and address the Hamiltonian for the OAM entanglement transfer. The key part will be the introduction of an adiabatic Hamiltonian involving Raman processes where photonic OAM quanta are used in order to put a single BEC into a vortex state with non-zero winding number. Sec.~\ref{Transfer} represents the core of our proposal by providing a clear physical picture of the process discussed above and assessing the quantitative performance of the light-to-BEC entanglement transfer. By starting from an OAM-carrying photonic state, we show that significant vortex-vortex entanglement is established by our scheme for experimentally realistic parameters. Sec.~\ref{reveal} describes realistic techniques for the inference of such quantum correlations based on the reverse of the map addressed here. Finally, in Sec.~\ref{conclusioni} we summarize our findings and open up perspectives for further development. A few technical steps which complement the quantitative analysis presented in the body of the paper are contained in Appendix A.

\begin{figure}[ht]s
\includegraphics[scale=0.35]{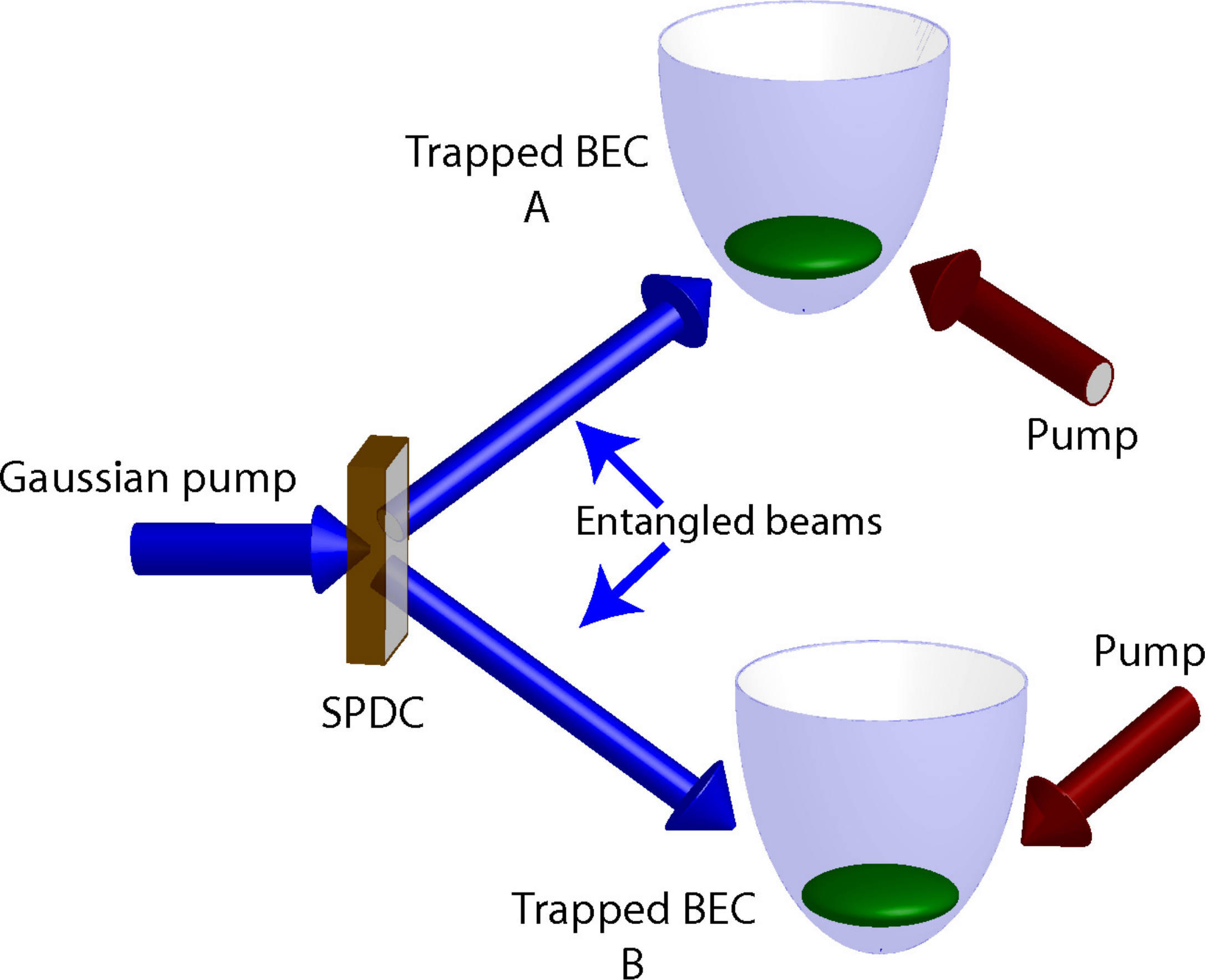}
\caption{(Color online). Sketch of the proposed setup. An OAM-entangled two-photon state is produced by spontaneus parametric down conversion (SPDC) of a Gaussian pump. Each output mode interacts with a respective trapped Bose-Einstein condensate (BEC), which is also pumped by an intense field with no OAM. The local matter-light interaction transfers the OAM entanglement from the field modes to the condensates rotational degree of freedom.}
 \label{fig:model}
\end{figure} 

\section{The model}
\label{modello}

Let us start by presenting the model used in order to describe the light-to-BEC transfer of entanglement. We shall see that a key point in this mechanism resides in the collective coupling of the atoms belonging to one of the BECs to the respective light field. Together with the indistinguishability of the resource photons, this permits us to entangle the two BECs. We consider two spatially separated and trapped BECs, each with $N^{I}_0$ ($I=A,B$) ${}^{87}$Rb atoms and let each of them interact with one of the field modes of an OAM-entangled two-photon state (see the sketch in Fig.~\ref{fig:model}). Such a photonic resource can be produced, for instance, by type-I parametric down conversion of a Gaussian laser beam
% at a wavelength $\lambda_{G}=35$nm which return photons at $\lambda_{G}=702$ nm. 
%In Ref.~\cite{entphot} it was shown that parametric down conversion 
which is an OAM-preserving process: \textcolor{black}{ the sum of the OAM carried by the entangled signal and idler mode produced by a laser-pumped non-linear crystal equals the OAM initially carried by the pump~\cite{entphot}. In this paper we shall consider a two-photon state produced by SPDC of a laser beam with no-OAM ({\it i.e.} prepared in a Gaussian spatial mode).} This implies that the output modes, here labeled as $\alpha$ and $\beta$, carry opposite OAM and enter the state
\begin{equation}
\label{eq:entphot}
\ket{\Phi}_{\alpha\beta}{\simeq}C_{0}\ket{1_{0}}_{\alpha}\ket{1_{0}}_{\beta}+C_{1,-1}\ket{1_{1}}_{\alpha}\ket{1_{-1}}_{\beta}+C_{-1,1}\ket{1_{-1}}_{\alpha}\ket{1_{1}}_{\beta}.
\end{equation}
Here, $\ket{n_{k}}_{\alpha}$ indicates an $n$-photon state populating mode $\alpha$ and carrying OAM equal to $\hbar{k}$. Moreover,  the presence of $\ket{1_{0}}_{\alpha}\ket{1_{0}}_{\beta}$ in $\ket{\Phi}_{\alpha\beta}$ accounts for a residual component with no OAM and $|C_{0}|^2+|C_{1,-1}|^2+|C_{-1,1}|^2\simeq1$ (having neglected components with higher OAM, which are very weakly populated at moderate pump intensities). The main idea behind our proposal is that the arrangement of a locally-assisted OAM-transfer from a light mode to the respective BEC would also allow for the transfer of quantum correlations, therefore constructing an effective entangled channel involving remote matter systems.

\begin{figure}[t]
\includegraphics[scale=1]{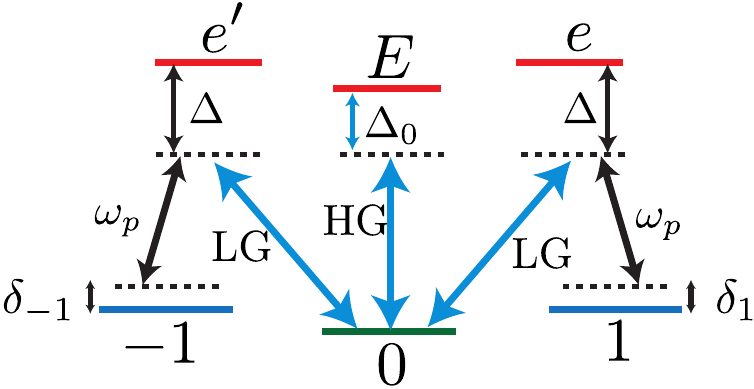}
\caption{(Color online). Six-level configuration for OAM transfer. We show a schematic representation of the relevant energy levels of a single ${}^{87}$Rb atom interacting with Laguerre-Gauss (LG) driving fields and classical Gaussian pumps with frequency $\omega_p$. The ground-state triplet comprises states having angular momentum $0$ and $\pm\hbar{l}$. The excited-state triplet is adiabatically eliminated from the dynamics by a double off-resonant Raman transition with the one-photon detunings $\Delta$ and $\Delta_0$. Two-photon detunings $\delta_{\pm{l}}$ are also introduced for the stabilization of the entanglement-transfer process. The component in Eq.~{\ref{eq:entphot}} carrying zero OAM [being in a Hermite-Gauss (HG) spatial mode] drives off-resonantly the $\ket{0}\leftrightarrow\ket{E}$ transition.}
 \label{fig:Mconfig}
\end{figure}

The basic building block for the transfer is an off-resonant double Raman scattering process. We consider each individual atom as a six-level system, shown in Fig.~\ref{fig:Mconfig}. The energy scheme comprises a ground-state triplet made out of a non-rotating state $\ket{0}$ and two other states, indicated as $\ket{\pm{1}}$, having angular momentum $\pm\hbar$. The elements of excited-state triplet $\ket{e}$ and $\ket{e^{'}}$ are linked to $\ket{\pm{1}}$ by two classical pumps, while the field modes in $\ket{\Phi}_{\alpha\beta}$ drive the $\ket{0}\leftrightarrow\ket{E,e,e'}$ transitions. In what follows, $\Delta$ and $\Delta_0$ indicate the one-photon Raman detunings \textcolor{black}{which are set by appropriate chosing the frequencies of the driving fields}. The classical pumps are taken to have a Gaussian spatial profile \textcolor{black}{so that photons scattered in the $\ket{e}\leftrightarrow\ket{1}$ and $\ket{e'}\leftrightarrow\ket{-1}$ transitions carry no OAM.} Together with the conditions on the OAM properties of Eq.~(\ref{eq:entphot}), this ensures that an atom undergoing the two-photon Raman transition \textcolor{black}{from state $\ket{0}$ to $\ket{\pm{1}}$ (as shown in Fig.~\ref{fig:Mconfig})} acquires an OAM exactly equal to $\pm \hbar$.

We now introduce the second-quantized matter field operators $ \hat{\psi}_{I,j}(\mathbf{r})$ obeying the bosonic commutation rules $ [\hat{\psi}_{I,i}(\mathbf{r}),\hat{\psi}_{J,j}^{\dagger}(\mathbf{r}^{'})]=\delta_{I,J}\delta_{i,j}\delta (\mathbf{r}-\mathbf{r}^{'})$. Here $I,J=A,B$ are labels for the BECs while $i,j=0,\pm{l},E,e,e^{'}$ refer to the atomic states. As for the photonic part of our system, standard  commutation relations $[\hat{c}_{n_{I},k}, \hat{c}_{n_{J},k^{'}}^{\dagger}]=\delta_{n_{I},n_{J}}\delta_{k,k^{'}}$ involving the creation (annihilation) operator $\hat{c}^\dag_{n_I,k}$ ( $\hat{c}_{n_I,k}$) hold. Here $n_{A}=\alpha$ ($n_{B}=\beta$) refers to the photonic mode $\alpha$ ($\beta$) that interacts with condensate $A$ ($B$). Finally, $k,k'=\pm{l},0$ refers to the OAM degree of freedom of the photon.

Besides the term describing the energy of the free photonic fields, the Hamiltonian of the system consists of the following four terms
\begin{equation}
\label{eq:fullham}
\hat{H} = \hat{H}_{a}+\hat{H}_{aa}+\hat{H}_{ad}+\hat{H}_{ap},
\end{equation}
which we now address in detail. The first two terms describe the properties of the trapped BECs and are given by
\begin{equation}
\hat{H}_{a}=\sum_{I,j} \int_{{\cal V}_{I}} d\mathbf{r}\; \hat{\psi}_{I,j}^{\dagger}(\mathbf{r})  \left(-\frac{\hbar^{2}}{2m}\nabla_{I}^{2}+V_{I,j}(\mathbf{r})\right) \hat{\psi}_{I,j}(\mathbf{r}),
\end{equation}
\begin{equation}
\label{eq:Haa}
\hat{H}_{aa}=\frac{1}{2}\sum_{I,j,j^{'}}\eta_{j,j^{'}}^{I} \int_{{\cal V}_{I}}d \mathbf{r} \; \hat{\psi}_{I,j}^{\dagger}(\mathbf{r})\; \hat{\psi}_{I,j^{'}}^{\dagger}(\mathbf{r})\; \hat{\psi}_{I,j^{'}}(\mathbf{r}) \;\hat{\psi}_{I,j}(\mathbf{r}).
\end{equation}
In all the above, ${\cal V}_I$ is the quantization volume for the light-matter interaction involving condensate $I$.
The $V_{I,j}(\mathbf{r})$s are the state-dependent atomic trapping potentials, $\eta_{j,j^{'}}^{I}=({4\pi\hbar^{2}}/{m})a_{j,j^{'}}^{I}$ accounts for the collisional energy between two atoms (of mass $m$) in states $j$ and $j^{'}$ and $a_{j,j^{'}}^{I}$ is the corresponding s-wave scattering length. As it will be clarified later on, the excited triplet $\{E,e,e'\}$ can be adiabatically eliminated from the dynamics of the atomic system. Therefore, by assuming no initial population of these states, we can simplify our treatment and take $a_{j,E}^{I}=a_{j,e}^{I}=a_{j,e^{'}}^{I}=0$ for $j=0,\pm{l}$. The third term in Eq.~(\ref{eq:fullham}) describes the interaction between BEC $I$ and the quantized field mode $n_I$ and can be written as
\begin{equation}
\begin{aligned}
&\hat{H}_{ad}=\sum_{I=A,B}\left(\chi_{I,0}\;\hat{c}_{n_{I},0} \int_{{\cal V}_{I}} d\mathbf{r}\,\hat{\psi}_{I,E}^{\dagger}(\mathbf{r})\hat{\psi}_{I,0}(\mathbf{r})\mathscr{A}_{n_{I},0}(\mathbf{r})\right.\\
&+\chi_{I,l}\,\hat{c}_{n_{I},l} \int_{{\cal V}_{I}} d\mathbf{r} \; \hat{\psi}_{I,e}^{\dagger}(\mathbf{r})\hat{\psi}_{I,0}(\mathbf{r})\mathscr{A}_{I,l}(\mathbf{r})\\
&\left.+\chi_{I,-l}\,\hat{c}_{n_{I},-l} \int_{{\cal V}_{I}} d\mathbf{r} \; \hat{\psi}_{I,e^{'}}^{\dagger}(\mathbf{r})\hat{\psi}_{I,0}(\mathbf{r})\mathscr{A}_{I,-l}(\mathbf{ r})\right)+h.c.
\end{aligned}
\end{equation} 

The coefficients $\chi_{I,k}~(k=0,\pm{l})$
%=C_{F,F^{'}}d_{F.D.}$s 
are the effective dipole moments associated with the transitions depicted in Fig.~\ref{fig:Mconfig}. %between the two hyperfine levels $F$ and $F^{'}$ considered and $d_{F.D.}$ is the values of the dipole moment for a far-detuned transition.
%We consider the same dipole moment for all the excited states because as we shall see an atom in whatever of these states has the same internal quantum numbers so the dipole moment has to be the same.\\
The functions $\mathscr{A}_{n_{I},k}(\mathbf{r})$ describe the spatial shape of the states entering $\ket{\Phi}_{\alpha\beta}$ and we choose them to be
\begin{equation}
\begin{aligned}
\mathscr{A}_{n_{I},0}(\mathbf{r})&=i\sqrt{\frac{\hbar\omega_{0}}{2\epsilon_{0}{\cal V}_{I}}}e^{i k_{z} z}e^{-\frac{r^2}{{\cal W}^2}},\\
\mathscr{A}_{n_{I},\pm l}(\mathbf{r})&=i\sqrt{\frac{\hbar\omega_{\pm l}}{2\epsilon_{0}{\cal V}_{I}}}\left(\frac{\sqrt{2}r}{\cal W}\right)^{\modulo{l}}e^{\pm{i}l\phi}e^{i k_{z} z}e^{-\frac{r^2}{{\cal W}^2}},
\end{aligned}
\end{equation}
here $\epsilon_{0}$ is the vacuum permeability.
While the first of these equations refers to a field mode having a Gaussian spatial profile, the second describes OAM-carrying Laguerre-Gauss beams. We assume that the beam-waist ${\cal W}$ is larger than any linear dimension of the BECs so that the Gaussian part of the function reduces to a constant. This approximation also ensures the collective nature of the interaction between the fields and the atoms belonging to a given BEC. The last term in $\hat{H}$ describes the coupling between the classical pumps and the BECs 
\begin{equation}
\begin{aligned}
\hat{H}_{ap}&=\sum_{I=A,B}\left(\hbar\Omega_{I} e^{-i \omega_{I}t}\!\int_{{\cal V}_{I}} d\mathbf{r}\,\hat{\psi}_{I,e}^{\dagger} (\mathbf{r}) \hat{\psi}_{I,l}(\mathbf{r})e^{i \mathbf{k}_{I} \cdot \mathbf{r}}\right.\\
&\left.+\hbar\Omega_{I} e^{-i \omega_{I}t}\!\int_{{\cal V}_{I}} d\mathbf{r} \, \hat{\psi}_{I,e^{'}}^{\dagger} (\mathbf{r}) \hat{\psi}_{I,-l}(\mathbf{r}) e^{i \mathbf{k}_{I}\cdot \mathbf{r}}\right)+h.c.
\end{aligned}
\end{equation}
The coefficients $\Omega_{I}$ are the Rabi frequencies for the matter-pump interactions. 
%and $E_{I}$ is the magnitude of the electrical field along the direction of the dipole moment $\mathbf{d}$\\
Moreover, we initially prepare each condensate in the atomic state $\ket{0}$ by means of optical pumping techniques, for instance, in such a way that the excited triplet can be considered as empty.
We now proceed with the adiabatic elimination of the excited states under the assumption of large single-photon detunings $\Delta$ and $\Delta_{0}$ (with respect to the typical coupling rates entering $\hat{H}_{aa}$, $\hat{H}_{ad}$ and $\hat{H}_{ap}$).  We take the time evolution due to the fields as faster than the center of mass motion of the atom when in one of the excited levels, so that we can neglect the free atomic Hamiltonian for these states. We move to a proper rotating frame where we redefine the excited-state (ground-state) matter field operators as
% with respect to the at the frequency of the Further we suppose that the time evolution of the operators $\hat{\psi}_{I,e}$, $\hat{\psi}_{I,e^{'}}$ and $\hat{\psi}_{I,E}$ is only determined by the drive field. Defining 
$\hat{\psi}_{I,e}= \hat{\tilde \psi}_{I,e} e^{-i \omega_{l}t}$, $\hat{\psi}_{I,e^{'}}=\hat{\tilde \psi}_{I,e^{'}} e^{-i \omega_{-l}t}$ and $\hat{\psi}_{I,E}= \hat{\tilde \psi}_{I,E} e^{-i \omega_{0}t}$
($\hat{\tilde \psi}_{I,k}=e^{-i \omega_{I}t}\hat{\psi}_{I,k}$) and the photonic operators as 
$\hat{\tilde c}_{I,k}(t)=e^{i \omega_{I}t}\hat{c}_{I,k}$ (with $k=0,\pm{l}$). Following the works by Marzlin {et al.}~\cite{marzlin} and Kapale and Dowling~\cite{dow}, we explicitly allow for two-photon Raman detunings $\delta_{\pm{l}}(t)=\omega_{l}-\omega_{p}(t)-\tilde \omega_{\pm{l}}$ (see Fig.~\ref{fig:Mconfig}), which help in the stabilization of the transfer process (see also Ref.~\cite{than}). Here $\hbar\tilde \omega_{\pm l}$ are the actual energies of the rotating atomic states. In fact, one can intuitively understand the necessity for a time-dependent two-photon detuning as a result of the adiabatic elimination of the excited triplet and the existence of inter-atomic collisions, which change the energies of the atom in time. Therefore, in order to achieve efficient transfer of OAM entanglement, we need a ``chirped" frequency of the pump fields that allows to track and compensate the change of the energy levels. The technical details behind the adiabatic elimination sketched here are presented in Appendix A.

We thus arrive at the effective Hamiltonian $\hat{H}_{eff} = \hat{\tilde H}_{a}+ \hat{H}_{aa}+ \hat{\tilde H}_{int}$ for the description of the adiabatic interaction between light and BECs, where
%\begin{widetext}
\begin{equation}
\label{eq:effham1}
\hat{\tilde H}_{a}\!=\!\sum_{I,j} \int_{{\cal V}_{I}}\!d\mathbf{r}\,\hat{\tilde \psi}_{I,j}^{\dagger}(\mathbf{r})\!\left(\!-\frac{\hbar^{2}}{2m}\nabla_{I}^{2}+V_{I,j}(\mathbf{r}) + \epsilon_{j}(t)\!\right)\!\hat{\tilde \psi}_{I,j}(\mathbf{r}),
\end{equation}
%\begin{equation}
%\label{eq:effham2}
%\hat{\tilde{H}}_{aa}=\frac{1}{2} \sum_{I,j,j^{'}}\eta_{j,j^{'}}^{I} \int_{{\cal V}_{I}}d \mathbf{r} \hat{\psi}_{I,j}^{\dagger}(\mathbf{r}) \hat{\psi}_{I,j^{'}}^{\dagger}(\mathbf{r}) \hat{\psi}_{I,j^{'}}(\mathbf{r}) \hat{\psi}_{I,j}(\mathbf{r}),
%\end{equation}
\begin{equation}
\label{eq:effham3}
\begin{aligned}
&\hat{\tilde H}_{int}\!=\!\sum_{I,k}\left[\hat{\tilde c}_{n_{I},k}\!\int_{{\cal V}_{I}}\!d\mathbf{r}\mathscr{C}_{n_I,k}({\bf r})\hat{\tilde \psi}_{I,k}^{\dagger}(\mathbf{r})\hat{\tilde \psi}_{I,0}(\mathbf{r}) +h.c.\right.\\
&\left.+\hat{\tilde c}_{n_{I},k}^{\dagger}\hat{\tilde c}_{n_{I},k}\int_{{\cal V}_{I}} d \mathbf{r} \frac{\chi_{I,l}^{2}\modq{\mathscr{A}_{n_{I},k}(\mathbf{r})}}{\Delta_{I}}\hat{\tilde \psi}_{I,0}^{\dagger}(\mathbf{r})  \hat{\tilde \psi}_{I,0}(\mathbf{r})\right]\\
&+\sum_{I}\hat{c}_{n_{I},0}^{\dagger}\hat{\tilde c}_{n_{I},0} \int_{{\cal V}_{I}} d \mathbf{r} \frac{\chi_{I,0}^{2}\modq{\mathscr{A}_{n_{I},0}(\mathbf{r})}}{\Delta_{I,0}}\hat{\tilde \psi}_{I,0}^{\dagger}(\mathbf{r})  \hat{\tilde \psi}_{I,0}(\mathbf{r}).
\end{aligned}
\end{equation}
%\end{subequations}
%\end{widetext}
In Eq.~(\ref{eq:effham1}) $j=0,\pm{l}$ should be taken, while in Eq.~(\ref{eq:effham3}) it is $k=\pm{l}$. Moreover, we have introduced the coupling coefficient $\mathscr{C}_{n_I,k}({\bf r})=({\hbar\Omega_{I}^{*}\chi_{I,k}}/{\Delta_{I}})\mathscr{A}_{n_{I},k}(\mathbf{r})e^{-i\mathbf{k}_{I} \cdot \mathbf{r}}$ and the energies $\epsilon_{j}(t)$ such that $\epsilon_{0}(t)=\hbar\omega_{I}$ and $\epsilon_{\pm l}(t)=\hbar (\omega_{\pm l}-\delta_{\pm{l}}(t)-\tilde \omega_{\pm l})$. The term describing inter-particle collisions $\hat{{H}}_{aa}$ remains identical to Eq.~(\ref{eq:Haa}). The interpretation of the form taken by $\hat{H}_{eff}$ is straightforward. While Eq.~(\ref{eq:effham1}) describes the energy of non-interacting matter part of the system, modified by the introduction of $\delta_{\pm}$-related terms, Eq.~(\ref{eq:effham3}) accounts for the light-matter interaction and includes the dynamical a.c. Stark shift effect arising from the adiabatic elimination. In particular, the first term in $\hat{\tilde H}_{int}$ describes a three-mode interaction where photonic excitations are used in order to perform an atomic transition between ground-triplet states. This is the key to our analysis on OAM entanglement transfer and the starting point of our quantitative study.

%____________________________________________A quantitative example______________________________

\section{Three-mode expansion and light-induced transfer of OAM entanglement}
\label{Transfer}

\subsection{Bosonic-mode expansion}

In order to fix the ideas and discuss an experimentally relevant case, we consider the anisotropic harmonic trap potential $V_{I}(r,z)=(m/2)(\omega_{r}^{2}r^{2}+\omega_{z}^{2}z^{2})$ for each of the BECs used in our proposal.
%\begin{equation}
%\label{eq:potential}
%V_{I}(r,z)=\frac{1}{2}m(\omega_{r}^{2}r^{2}+\omega_{z}^{2}z^{2})
%\end{equation}
Here $\omega_{z}$ ($\omega_{r}$) is the frequency of the trap along the longitudinal (radial) direction. In order to be able to neglect any longitudinal excitations, we assume $\omega_{z}\gg\omega_{r}$, so that each BEC is confined in a pancake-like structure. In the limit where the inter-atomic collisions are very small, the cylindrical symmetry of the problem allows us to describe the centre-of-mass of one atom in a BEC by means of the set of eigenstates ($\theta$ is the angular coordinate of a cylindrical reference frame)
\begin{equation}
\label{eq:eigenstates}
\begin{aligned}
\phi_{I,0}(r,z)&=\frac{1}{\pi^{\frac{3}{4}}a_{r}\sqrt{a_{z}}}e^{-\frac{1}{2}\left(\frac{r^2}{a^2_{t}}+\frac{z^2}{a^2_{z}}\right)},\\
\phi_{I,\pm l}(r,z,\theta)&=\frac{1}{\sqrt{\modulo{l}!}a_{r}^{\modulo{l}}}r^{\modulo{l}}e^{\pm{i} l \theta}\phi_{I,0}(r,z)
\end{aligned}
\end{equation}
with associated eigenvalues $E_{l}={\hbar}\omega_{z}/2+\hbar(\modulo{l}+1)\omega_{r}$. Here $a_{z,r}$ are the characteristic lengths of the harmonic motion along the longitudinal and radial direction. The description provided by the eigenfunctions (\ref{eq:eigenstates}) remains valid under the assumption of dilute BECs, so that their ground states result from the simple tensor product of the single-particle states $\phi_{I,0}(r,z)$. This is a good approximation as long as $N^{I} a_{j,j^{'}}^{I}\ll a_{z}$~\cite{pitstring}, implying that the scattered part of the single-particle wavefunction contributes with only a small correction to the wave-function of the non-interacting case. In order to provide a better picture of the anticipated three-mode interaction depicted in Eq.~(\ref{eq:effham3}), we now define new bosonic operators $\hat{b}_{I,j}$ and $\hat{b}^\dag_{I,j}$ for the matter-like part of our system as (omitting the tilde from now on for readability) $\hat{\psi}_{I,j}(\mathbf{r})=\phi_{I,j}(\mathbf{r})\hat{b}_{I,j}$. By using the orthogonality of the $\phi_{I,j}({\bf r})$ and the commutation relations 
%Noting that $\int_{V_{I}}d \mathbf{r} \quad \phi_{I,j}^{*}(\mathbf{r})\phi_{I,j^{'}}(\mathbf{r})=\delta_{j,j^{'}}$ and because of the commutation relatioins of 
valid for $\hat{\psi}(\mathbf{r})$'s, it is straightforward to find that 
%we get the commutation relation for the new operators which are 
$[\hat{b}_{I,i},\hat{b}_{I,j}^{\dagger}]=\delta_{i,j}$.
Inserting these definitions into the effective Hamiltonian we obtain a much simplified and self-evident picture of the process through
\begin{equation}
\label{eq:bham}
\begin{aligned}
\hat{ H}_{a}&=\sum_{I,k}(E_{I,0}+\modulo{k}\hbar\omega_{r}+\epsilon_{k}(t))\hat{b}_{I,k}^{\dagger}\hat{b}_{I,k},\\
\hat{ H}_{aa}&=\frac{1}{2}\sum_{I,j,j^{'}}\xi_{j,j^{'}}^{I}\hat{b}_{I,j}^{\dagger}\hat{b}_{I,j^{'}}^{\dagger}\hat{b}_{I,j^{'}}\hat{b}_{I,j},\\
%\hat{\tilde H}_{d}&=-\hbar\omega_{I}\sum_{j} \hat{c}_{n_{I},j}^{\dagger}\hat{c}_{n_{I},j}\\
\hat{ H}_{int}&=\sum_{I,k=l,-l}\big(g_{I,k}\hat{b}_{I,k}^{\dagger}\hat{b}_{I,0}\hat{c}_{n_{I},k}+h.c.\big)\\
&+\hat{b}_{I,0}^{\dagger}\hat{b}_{I,0}\sum_{I,k}\rho_{I,k}\hat{c}_{n_{I},k}^{\dagger}\hat{c}_{n_{I},k},
\end{aligned}
\end{equation}
where each coefficient can be expressed in terms of the non-interacting wave-functions as shown in Table~\ref{table:coefficienti}.\\ 
\begin{table}[b]
\caption{Coupling rates in the effective Hamiltonian after the introduction of the effective matter-like bosonic operators [see Eqs.~(\ref{eq:bham})] and their expressions in terms of the non-interacting atomic wave-function for a pancake-like potential.}
\centering
\begin{tabular}{c c}
\hline
\hline
Coefficients & Corresponding expression \\
\hline
\hline
$E_{I,0}$ & $\int_{{\cal V}_{I}}d\mathbf{r}\phi_{I,0}^{*}(\mathbf{r})(-\frac{\hbar^{2}}{2m}\nabla_{I}^{2}+V_{I,j}(\mathbf{r}))\phi_{I,0}(\mathbf{r})$\\
$g_{I,k}$ & $\int_{{\cal V}_{I}}d \mathbf{r} \mathscr{C}_{n_{I},k}(\mathbf{r})\phi_{I,k}^{*}(\mathbf{r})\phi_{I,0}(\mathbf{r})$\\
$\rho_{I,0}$ & $\int_{{\cal V}_I} d \mathbf{r}\frac{\chi_{I,0}^{2}\modq{\mathscr{A}_{n_{I},0}(\mathbf{r})}}{\Delta_{I,0}} \modq{\phi_{I,0}(\mathbf{r})}$\\
$\rho_{I,l}$ & $\int_{{\cal V}_I} d \mathbf{r}\frac{\chi_{I,l}^{2}\modq{\mathscr{A}_{n_{I},l}(\mathbf{r})}}{\Delta_{I}} \modq{\phi_{I,0}(\mathbf{r})}$\\
$\xi_{j,j^{'}}^{I}$ & $\eta_{j,j^{'}}^{I}\int_{\cal V_{I}}d \mathbf{r}\modq{\phi_{I,j}(\mathbf{r})} \modq{\phi_{I,j^{'}}(\mathbf{r})}$\\
\hline
\hline
\end{tabular}
\label{table:coefficienti}
\end{table}
%\begin{eqnarray}
%&{}&\nonumber\\
%E_{I,0}&=&\int_{V_{I}}d\mathbf{r} \quad \phi_{I,0}^{*}(\mathbf{r})  (-\frac{\hbar^{2}\nabla_{I}^{2}}{2m}+V_{I,j}(\mathbf{r}))\phi_{I,0}(\mathbf{r})\nonumber\\
%&{}&\nonumber\\
%g_{I,m}&=&\frac{\hbar\Omega_{I}^{*}\chi_{I,m}}{\Delta_{I}}\int_{V_{I}}d \mathbf{r} \quad \mathscr{A}_{n_{I},m}(\mathbf{r})e^{-\imath\mathbf{k}_{I} \cdot \mathbf{r}}\phi_{I,m}^{*}(\mathbf{r})\phi_{I,0}(\mathbf{r})\nonumber\\
%&{}&\nonumber\\
%\rho_{I,0}&=& \int_{V} d \mathbf{r} \quad \frac{\chi_{I,0}^{2}\modq{\mathscr{A}_{n_{I},0}(\mathbf{r})}}{\Delta_{I,0}} \modq{\phi_{I,0}(\mathbf{r})}\nonumber\\
%&{}&\nonumber\\
%\rho_{I,l}&=& \int_{V} d \mathbf{r} \quad \frac{\chi_{I,l}^{2}\modq{\mathscr{A}_{n_{I},l}(\mathbf{r})}}{\Delta_{I}} \modq{\phi_{I,0}(\mathbf{r})}\nonumber\\
%&{}&\nonumber\\
%\xi_{j,j^{'}}^{I}&=&\eta_{j,j^{'}}^{I}\int_{V_{I}}d \mathbf{r} \quad \modq{\phi_{I,j}(\mathbf{r})} \modq{\phi_{I,j^{'}}(\mathbf{r})}\nonumber\\
%&{}&\nonumber
%\end{eqnarray}
The effect of the light-matter coupling is now manifest: besides the a.c. Stark shifts proportional to $\rho_{I,j}$, $\hat{H}_{int}$ consists of a scattering process at a rate $g_{I,k}$ where the annihilation (creation) of a photon of angular momentum $k$ is accompanied by the Raman transition $\ket{0}_{I}\rightarrow\ket{k}_{I}$ ($\ket{k}_{I}\rightarrow\ket{0}_{I}$). Such a mechanism, which would determine a perfect transfer of OAM from the light resource to the BECs, is {\it disturbed} by the inter-atomic collisions in $\hat{H}_{aa}$ and should also take into account the modifications induced by the dynamical shifts in $\hat{H}_{a,int}$. The creation of inter-BEC OAM entanglement is thus a trade-off between these various processes. The task of the next Subsection is precisely the quantitative assessment of such a trade-off. It is worth remarking here that, in virtue of the definitions of $\epsilon_k(t)$ the term $\modulo{k}\hbar\omega_{r}+\epsilon_{k}(t)$ appearing in the energy of the rotating states with $j=\pm{l}$ is explicitly dependent on the two-photon Raman detunings $\delta_k(t)$ and takes the form  
% ,  and remembering the expression for $\epsilon_{j}$ this term tour out to be 
$\hbar(\omega_{k}-\delta_{k}-\tilde\omega_{k}+\modulo{k}\omega_{r})$. When the interactions considered in our scheme are included, the energy levels are shifted so that the shift $\tilde\omega_{k}-\modulo{k}\omega_{r}\neq0$ is in general non-zero and, possibly, time-dependent. In what follows, we  shall assume that such shift occurs linearly in time.

\subsection{Entanglement transfer process}

%We can now study the time evolution of our system under the action of the Hamiltonian (\ref{eq:bham}).\\
As discussed in Sec.~\ref{modello}, we assume an initial preparation where the atomic excited triplet is empty and
%, by means of optical pumping techniques, 
all the atoms in each BEC populate $\ket{0}_I$. We indicate such a collective atomic state as $\ket{N^{I}_0}_I$, which condenses information on the population of level $\ket{0}_I$ and $\ket{\pm{1}}_{I}$. On the other hand, the OAM-entangled photonic resource is taken as prepared in the experimentally realistic state $\ket{\Phi_{Z}}_{\alpha\beta}=({1}/{\sqrt{3}})(\ket{1_{0},1_{0}}_{\alpha\beta}+\ket{1_{1},1_{-1}}_{\alpha\beta}+\ket{1_{-1},1_{1}}_{\alpha\beta})$.
\textcolor{black}{(Here we put the label $Z$ which refers to the state obtained by Zeilinger and cooworkers)} In Refs.~\cite{entphot,entphot2} it was shown that two-photon multidimensional OAM-entangled states generated by means of SPDC can be effectively distilled into states very close to $\ket{\Phi_Z}_{\alpha\beta}$, thus making the contributions coming from states having higher OAM negligible~\cite{commentoT}. This effectively makes the Hilbert space spanned by photonic OAM states isomorphic to that of a spin$-1$ particle, or {\it qutrit}, so that $\ket{\Phi_Z}_{\alpha\beta}$ describes a maximally entangled two-qutrit state. The initial state is thus taken to be $\ket{\Psi(0)}_{AB\alpha\beta}=\ket{N_{0}^{A},N_{0}^{B}}_{AB}\ket{\Phi_{Z}}_{\alpha\beta}$, whose dynamics under Eqs.~(\ref{eq:bham}) is now evaluated in a rotating frame defined according to Eq.~(A-2). 
%It is matter of a straightforward calculation to recognize that 

\begin{figure}[t]
\includegraphics[scale=0.95]{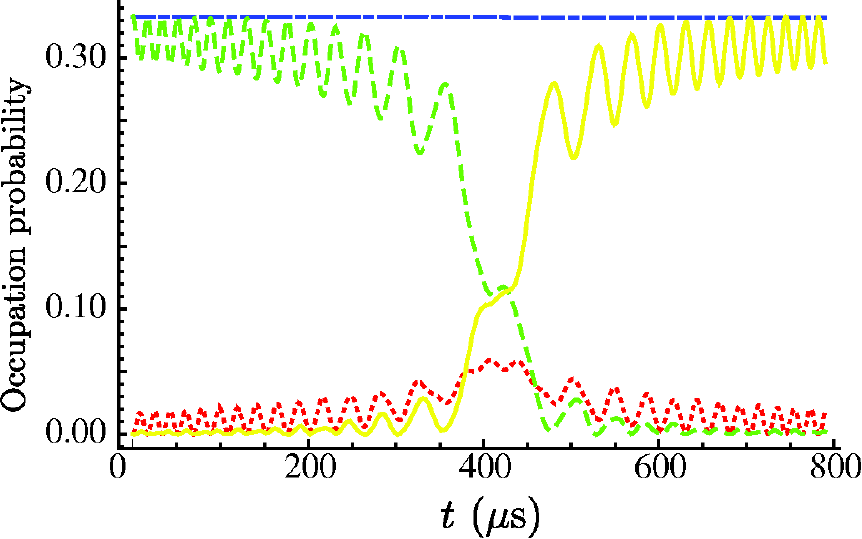}
\caption{(Color online).Evolution of the state probabilities $\modq{f_{i}(t)}$ against the interaction time $t$ (in $\mu$s) for $N^I_0=10^5$ atoms per condensate. Here $f_{j,j^{'}}^{I}=5 nm$, $\Omega_{I}=27.5$kHz, $\chi_{I,\pm{l}}=1.18$kHz, $\Delta=90$kHz and the two-photon detunings having the functional form $\delta_{\pm{l}}(t)=2\Omega_{I}(1-\Omega_{I}t/2)-\omega_t$. The trap frequencies are $\omega_t=70$Hz and $\omega_z=500$Hz. The photonic resource is tailored at a wavelength of $702$nm {at an angle of $4^{o}$ off the initial pumping gaussian beam's axis}}
\label{fig:prob105}
\end{figure}

%\begin{equation}
%\label{eq:schroe}
%\imath \hbar \frac{d}{dt}\ket{\Psi(t)}=(\hat{\tilde H}-\omega_{I}\hat{O})\ket{\Psi(t)}
%\end{equation}
%where the second term rises because of the transformations (\ref{eq:transf}).\\
\textcolor{black}{It is straightforward to verify that, when starting from $\ket{\Psi(0)}_{AB\alpha\beta}$ as given above, the evolved state $\ket{\Psi(t)}_{AB\alpha\beta}$ obtained upon use of the effective Hamiltonian in Eq.~(\ref{eq:bham}) lies entirely in a nine-dimensional sector of the Hilbert space spanned by the states}
\begin{equation}
\label{chiusi}
\begin{aligned}
&\ket{\Psi_{0}}=\ket{N_{0}^{A}}_A\ket{N_{0}^{B}}_{B}\ket{1_{0}}_{\alpha}\ket{1_{0}}_{\beta},\\
&\ket{\Psi_{1}}=\ket{N_{0}^{A}}_A\ket{N_{0}^{B}}_B\ket{1_{1}}_{\alpha}\ket{1_{-1}}_{\beta},\\
&\ket{\Psi_{2}}=\ket{N_{0}^{A}}_A\ket{N_{0}^{B}}_B\ket{1_{-1}}_{\alpha}\ket{1_{1}}_{\beta},\\
&\ket{\Psi_{3}}=\ket{N_{0}^{A}-1,1_{1}}_A\ket{N_{0}^{B}}_B\ket{0}_{\alpha}\ket{1_{-1}}_{\beta},\\
&\ket{\Psi_{4}}=\ket{N_{0}^{A}-1,1_{-1}}_A\ket{N_{0}^{B}}_B\ket{0}_{\alpha}\ket{1_{1}}_{\beta},\\
&\ket{\Psi_{5}}=\ket{N_{0}^{A}}_A\ket{N_{0}^{B}-1,1_{1}}_B\ket{1_{-1}}_{\alpha}\ket{0}_{\beta},\\
&\ket{\Psi_{6}}=\ket{N_{0}^{A}}_A\ket{N_{0}^{B}-1,1_{-1}}_B\ket{1_{1}}_{\alpha}\ket{0}_{\beta},\\
&\ket{\Psi_{7}}=\ket{N_{0}^{A}-1,1_{1}}_A\ket{N_{0}^{B}-1,1_{-1}}_B\ket{0}_{\alpha}\ket{0}_{\beta},\\
&\ket{\Psi_{8}}=\ket{N_{0}^{A}-1,1_{-1}}_A\ket{N_{0}^{B}-1,1_{1}}_B\ket{0}_{\alpha}\ket{0}_{\beta}.\\
\end{aligned}
\end{equation}
The notation used here is such that $\ket{N_{0}^{I}-s,s_{k}}_I$ indicates a state where $s$ atoms populate an atomic eigenstate of the angular momentum with eigenvalue $k\hbar$ while $N^I_0-s$ atoms populate the state $\ket{0}_I$ having zero OAM. \textcolor{black}{It is worth stressing that the number and structure of the states involved in the evolution of a given initial state strongly dependent on the total initial angular momentum carried by the latter. In fact, our effective Hamiltonian preserves the total light-matter OAM}. This property implies that the dynamically evolved state of the system should be written as
\begin{equation}
\label{eq:evol}
\ket{\Psi(t)}_{AB\alpha\beta}=\sum_{i=0}^{8}f_{i}(t)\ket{\Psi_{i}}
\end{equation}
with numerical coefficients $f_{i}(t)$ such that $\sum_{i}\modq{f_{i}(t)}\!=\!1$. The analytic solution of such a dynamics is a formidable problem and we thus resort to a numerical investigation in order to infer the behavior of $f_{i}(t)$'s. \textcolor{black}{To find the coefficients $\{f_{i}(t)\}$, we have numerically solved the Schr\"{o}dinger equation using the Hamiltonian in Eqs.~(\ref{eq:bham})~\cite{notaReferee}. We have explored a wide range of parameters, including the case where the system is symmetric under the exchange of the two BECs, finding qualitatively similar results. In the following, we concentrate on the symmetric case and use the parameters listed in the caption of Fig.~(\ref{fig:prob105}),} which shows that a complete transfer of OAM from the photonic state to the BECs is possible, in analogy with the semiclassical case approached in Refs.~\cite{marzlin,dow,than}. The (dashed) green curve shows the temporal dynamics of the probabilities $\modq{f_{1,2}(t)}$ whereas the (solid) yellow ones depict $\modq{f_{7,8}(t)}$ for a given set of the relevant physical parameters and a specific choice for the functional form of the chirped two-photon detunings. These two sets of probabilities are almost mutually mirror symmetric. Damped oscillations are superimposed to a monotonic behavior induced by the compensation arising from the chirped detunings (in our case $\delta_{\pm{l}}(t)=2\Omega_{I}(1-\Omega_{I}t/2)-\omega_t$).  The low-lying (dotted) red curve is for $\modq{f_{3,4,5,6}}$, whose corresponding states only marginally contribute to the evolution of the system. Finally, the (dot-dashed) horizontal blue line shows the probability $\modq{f_{0}(t)}$, which does not change in time as $\ket{\Psi_{0}}$ is an eigenstate of the effective Hamiltonian. As we shall see this gives a lower limit for the population of the ground state of the reduced density matrix for the two BECs. 
In fact a plot of the populations of the BECs reduced density matrix, $\rho_{AB}(t)=\text{Tr}_{\alpha\beta}(|{\Psi(t)}\rangle_{AB\alpha\beta}\langle{\Psi(t)}|)$ shows that the state $\ket{N_{0}^{A}}_A\ket{N_{0}^{B}}_B$ has always a finite occupation probability. This is due to the unavoidable presence of photons carrying no OAM that continuously project the two BECs onto their ground states. 

\begin{figure}[b]
\includegraphics[scale=1]{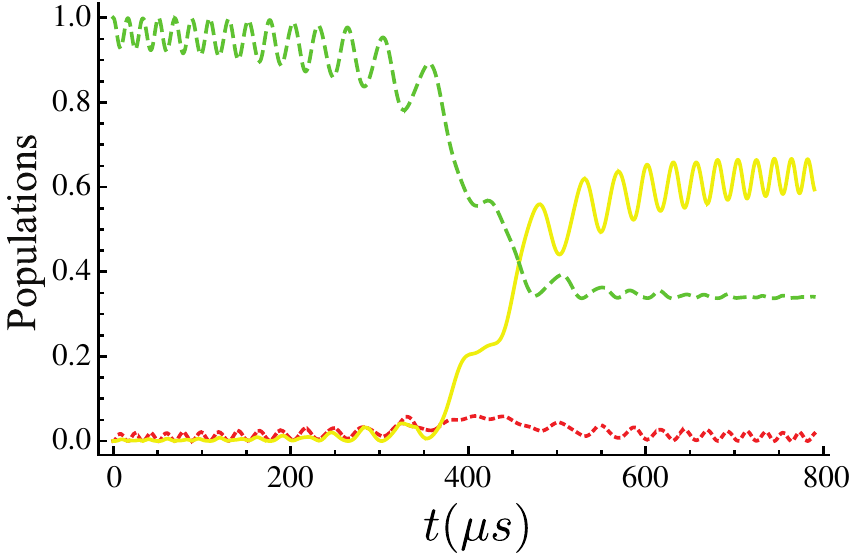}
\caption{(Color online). Time behavior of the populations of the atomic states in $\rho_{AB}(t)=\text{Tr}_{\alpha\beta}(|{\Psi(t)}\rangle_{AB\alpha\beta}\langle{\Psi(t)}|)$. The same parameters as in Fig.~\ref{fig:prob105} have been used here. The (dashed) green curve is for $\ket{N^A_0,N^B_0}_{AB}$, the (solid) yellow one is for state $\ket{B}$. The low-lying (dotted) red curves represent the probability that the remaining two-BEC basis states are excited. The incomplete population transfer from $\ket{0}_{I}$ to $\ket{\pm{l}}_{I}$ ($I=A,B$) is due to the zero-OAM terms in $\ket{\Phi_Z}_{\alpha\beta}$. }
\label{fig:pop105}
\end{figure}

The (solid) yellow curve in Fig.~\ref{fig:pop105} shows the population of the state 
%\begin{equation}
%\label{eq:adstates}
%\begin{aligned}
$\ket{B}\!=\!(\ket{N_{0}^{A}\!-\!1,1_{1}}\ket{N_{0}^{B}\!-\!1,1_{-1}}\!+\!\ket{N_{0}^{A}\!-\!1,1_{-1}}\ket{N_{0}^{B}\!-\!1,1_{1}})/\sqrt{2}$
%&\ket{{D}}\!=\!\frac{\ket{N_{0}^{A}\!-\!1,1_{1}}\ket{N_{0}^{B}\!-\!1,1_{-1}}\!-\!\ket{N_{0}^{A}\!-\!1,1_{-1}}\ket{N_{0}^{B}\!-\!1,1_{1}}}{\sqrt{2}},
%\end{aligned}
%\end{equation}
(we have omitted the BEC label as no ambiguity exists) which shows OAM entanglement between the BECs. Clearly, the OAM entanglement transfer generates quite a large component of $\ket{B}$ in the reduced two-BEC state. This arises from the OAM-carrying components in the photonic resource and the efficiency of the population-transfer process. %. At the same time, as the dark state $\ket{D}$ is decoupled from the photonic field, it is obvious that the corresponding population remains flat in time. 
The introduction of this state allows us to draw a clear and compact picture of the asymptotic form of the map $\hat{\cal M}_{t}$ transforming photonic OAM entanglement into matter-like one via bi-local far off-resonant double Raman coupling. By neglecting the very small components associated with the remaining excited two BEC states ((dotted) red line in Fig.~\ref{fig:pop105}) and collecting the remaining terms into a diagonal density matrix, this is approximately given by 
\begin{equation}
\label{eq:finale}
\begin{aligned}
&\lim_{t\rightarrow\infty}\hat{\cal M}_{t}(\ket{N^A_0,N^B_0}_{AB}\bra{N^A_0,N^B_0})\\
&\simeq\frac{1}{3}(2\ket{B}_{AB}\langle{B}|+\ket{N^A_0,N^B_0}_{AB}\bra{N^A_0,N^B_0}),
\end{aligned}
\end{equation} 
%where $\varphi(t)$ is a dynamical parameter determined by the solution o the BECs' evolution. 
Such a formal asymptotic map also explains that we can formally infer the properties of the reduced two-vortex density matrix by treating it as the state of two (in general entangled) qutrits. 

\begin{figure}[b]
\includegraphics[scale=1]{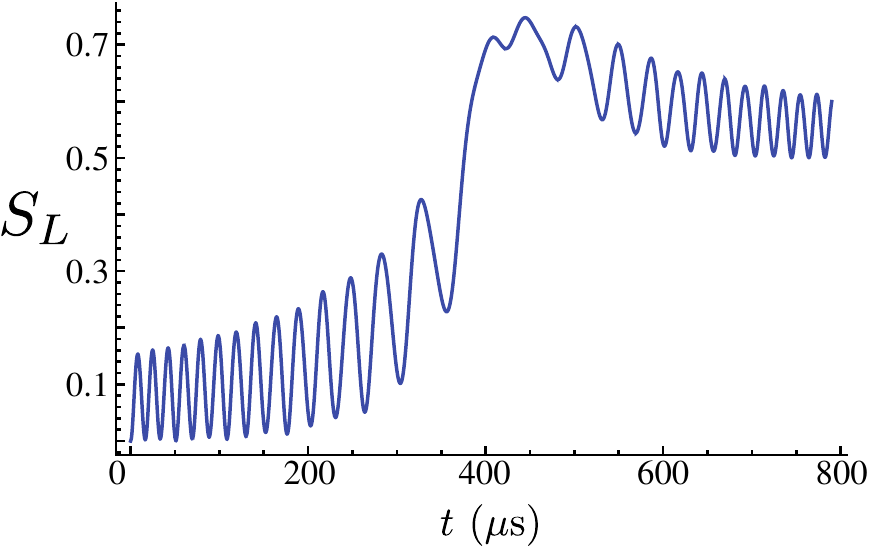}
\caption{(Color online). Linear entropy $S_{L}(\rho_{AB})$ against time $t$ (units of $\mu$s). The same parameters as in Fig.~\ref{fig:prob105} have been used here.}
\label{fig:pur105}
\end{figure}

\subsection{Assessment of entanglement}
We are now in a position to quantitatively estimate the amount of vortex entanglement set between the BECs. In order to tackle this point, our approach will be twofold. First, we study the time evolution of the {\it linearized entropy} $S_{L}(\rho_{AB}(t))=(9/8)[1-{\text{Tr}}(\rho_{AB}(t))]$~\cite{linent} of the BEC density matrix. As $S_{L}$ is a good measure of the purity of a state (it achieves $0$ for perfectly pure states and $1$ for statistical mixtures), this will give us an indication of the residual entanglement set between the photonic and matter-like part of the system: as the dynamics set by $\hat{H}_{eff}$ is unitary, the entanglement initially present in the photonic state has to be conserved when the whole state of the matter-light system is considered. Needless to say, such entanglement can be transferred from the photonic subsystem to the atomic one and/or vice versa and a transient can well exist where the two are almost separable. Such a situation would  be witnessed by a small value of $S_L$ and correspond to either large or small values of BEC entanglement, with minimal photons-atoms quantum correlations. We have determined the form of $S_L$ as a function of time, which is shown in Fig.~\ref{fig:pur105}. As expected, at exactly the time when the populations of the OAM carrying atomic system become non-zero, the linear entropy changes its behavior, signaling a maximum of mixedness of the light-matter state. This simply implies that for $t\in[400,500]\mu$s the two subsystems are correlated in a nonclassical sense. If time increases further, $S_{L}$ evidently decreases, witnessing a reduction in the light-matter entanglement. Because of the conservation of entanglement discussed above, this is the region we are interested in, as it could well be the case that in this long-time window significant inter-vortex entanglement is set at the expenses of the initial all-optical one and the transient matter-light correlations highlighted here.

\begin{figure}[t]
\includegraphics[scale=1]{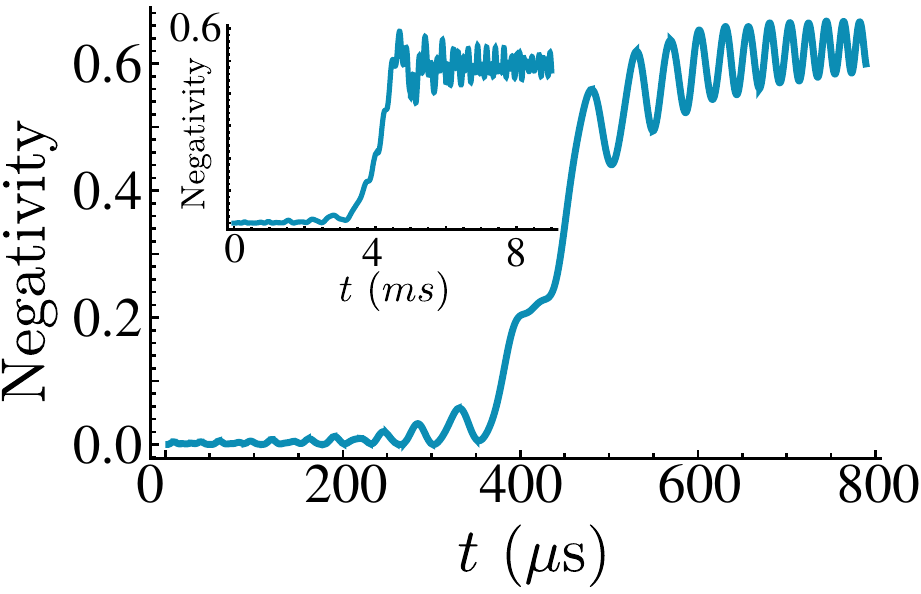}
\caption{(Color online). Negativity $N(\rho_{AB})$ against the interaction time $t$ (units of $\mu$s). The same parameters as in Fig.~\ref{fig:prob105} have been used here. We obtain a maximum of entanglement in correspondence of the range of decreasing trend of $S_L(\rho_{AB}(t))$, which witnesses a larger purity of the two-vortex state and smaller quantum correlations between light and matter. Inset: Negativity against $t$ for $N_{0}^{A}=10N_{0}^{B}=10^4$ and $\Omega_B=10\; \Omega_A= 27.5$ kHz. Other parameters as in Fig.~\ref{fig:prob105}.}
\label{fig:neg105}
\end{figure}

We quantitatively confirm our expectations by studying the {\it negativity}~\cite{peres,neg}, an entanglement measure based on the violation of the ``positivity of partial transposition" (PPT) criterion for separability of a state. Negativity is defined as~\cite{neg}
\begin{equation}
\label{nega}
N(\rho_{AB}(t))=-2\sum_k\lambda^-_k
%||\rho^{pt}_{AB}(t)||_{1}-1
\end{equation}
where $\lambda^-_k$ are the negative eigenvalues of the partially transposed density matrix with respect to one of the BEC systems.
%  and $||\cdot||_{1}$ stands for the trace norm (modulo $1$). 
The results are shown in Fig.~\ref{fig:neg105}.
As expected, the region of large inter-vortex entanglement corresponds to the range of interaction times where the linearized entropy decreases towards a steady-state value.
\textcolor{black}{We point out that the wavy behaviour of the curve in Fig.~(\ref{fig:neg105}), as well as in the other plots of the paper, is due to the inter-atomic scattering. This is confirmed by the the plot shown in the inset of the figure, where an asymmetric case has been studied.  The comparison between symmetric and asymmetric case shows that a mismatched number of atoms in the two BECs results in a change in the oscillatory behavior of the curve describing the time evolution of the transferred entanglement. On the other hand, mismatched Rabi frequencies only determine a change in the temporal scale of the entanglement dynamics. The degree of transferred entanglement is only mildly affected, which demonstrates the robustness of our protocol to such effects. The efficiency of our protocol does not depend on the assumption of symmetry under the exchange of the condensates and is retained in a wide range of the relevant parameters. Therefore, for the sake of convenience and without affecting the generality of our results, in what follows we restrict our attention to the symmetric case.}

As discussed above, quite a large value of entanglement is set between the vortex states of the two condensates, although a maximally entangled state [achieving $N(\rho_{AB})=1$] is not reached. We stress that this is not a limitation of our scheme but, on the contrary, an effect of the zero-OAM component in the photonic resource $\ket{\Phi_Z}_{\alpha\beta}$. Such a detrimental contribution can be removed from the BECs reduced state (and its properties) by resorting to an ``active" approach where, instead of discarding the state of light after the interaction with the condensates, we properly {\it post-select} its state. Upon inspection of $\ket{\Psi_{7,8}}$ in Eqs.~(\ref{chiusi}), it is straightforward to see that state $\ket{B}_{AB}$ is associated with modes $1$ and $2$ in the vacuum state. On the other hand, the {\it entanglement-spoiling} component $\ket{n^A_0,n^B_0}_{AB}$ would bring about photons in both the modes. It is therefore sufficient to use a standard Geiger-like avalanche photo-detector per mode, which discriminate the vacuum from the presence of any non-zero number of photons in a field, in order to operate the optimal post-selection of the BECs state: by registering no click at both the photo-detectors, we project the state of $A$ and $B$ onto the maximally entangled state $\ket{B}$. It is in fact worth stressing that, by 
effectively excluding the possibility that the atoms occupy state $\ket{0}_I$, the post-selection procedure further reduces the dimension of the relevant Hilbert space spanned by each vortex state to a bidimensional one, thus leaving us with two effective qubits.

\section{Detection of vortex entanglement}
\label{reveal}
In this Section we describe a method for the detection of the vortex entanglement created by the process above. Given the low-excitation level of our protocol, the usual  matter-wave interference is not helpful and we instead propose an approach based on the inversion of the process addressed here for light-to-BEC entanglement transfer. After generation of a two-vortex entangled state (as described in Sec.~\ref{Transfer}), the OAM-transferring interaction should be stopped. We thus assume that the pump fields have been turned off (or set far off-resonant with respect to the frequency of the transitions they guide) so as not to perturb the entangled states of the vortices.
The time-reversal nature of our protocol makes it intuitive to understand that, if we now reinstate such pumps, photons will be scattered into two Laguerre-Gauss modes at the frequency of the $\ket{0}_{I}\leftrightarrow\ket{e,e'}$ transition, thus {\it writing back} the two-vortex state onto light fields. One can then apply state-property reconstruction techniques, including testing Bell's inequality violation for bipartite states of effective three-level particles~\cite{3Bell}. However, the success of such tests is usually very sensitive to the form of the state under scrutiny and the level of non-ideality affecting it. 
We thus resort to a specific and quite promising way to infer the properties of the state we have generated based on Wigner function reconstruction, which is possible by using computer generated holograms~\cite{entphot} and homodyne-like measurements~\cite{cechi}.

\begin{figure}[b]
\includegraphics[scale=.4]{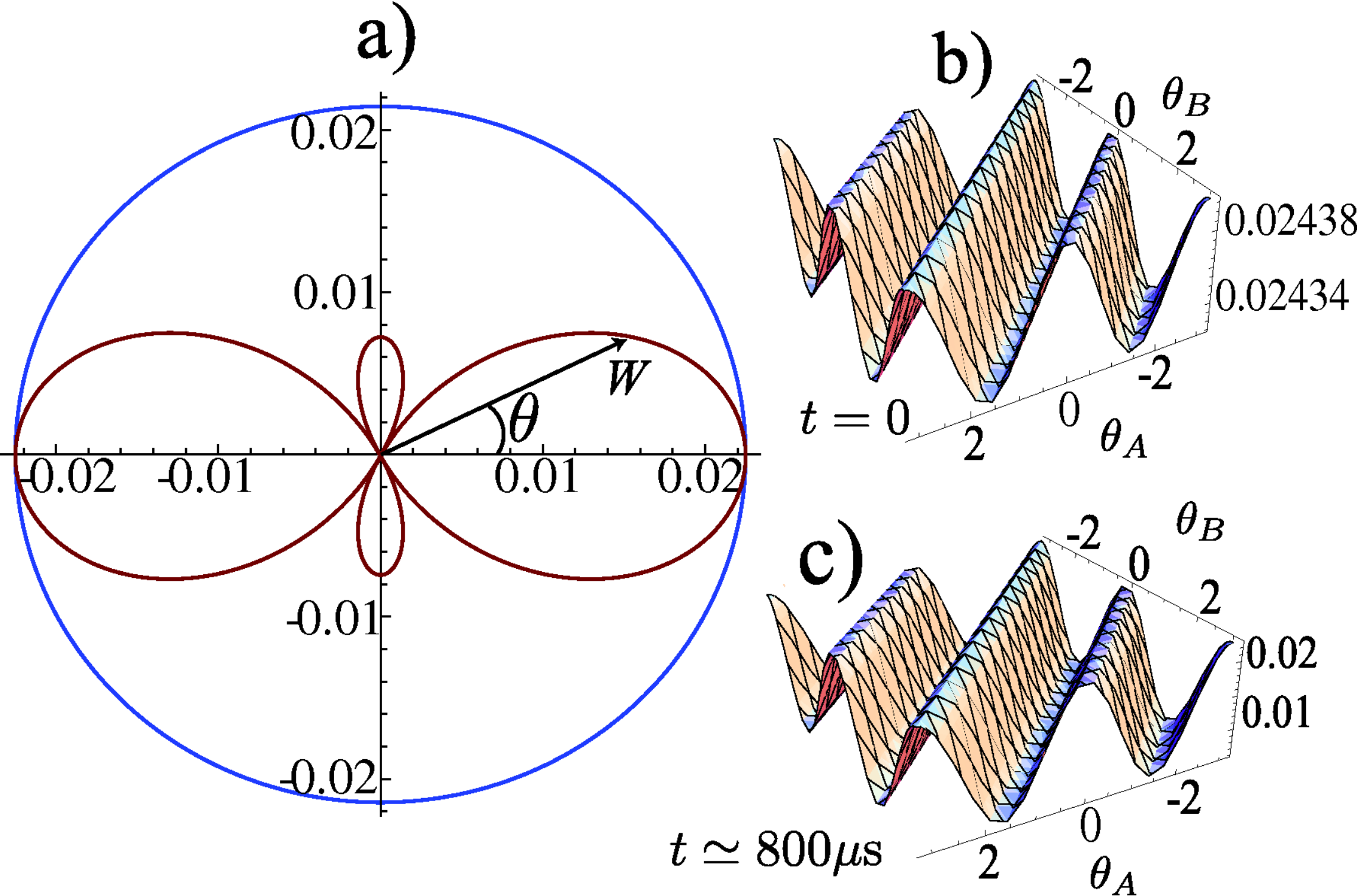}
\caption{(Color online). Analysis of the Wigner function for $\rho_{AB}(t)$. {\bf (a)}: The blue circle shows a projection the Wigner function on the unit circle for $t=0$, while the red butterfly structure is associated with $t\simeq{800}\mu$s, where OAM entanglement has been transferred. {\bf (b)} and {\bf (c)}: We plot $W(0,\theta_A,0,\theta_B)$ at the two instants of time considered for panel {\bf (a)}. Notice the different vertical-axis scales in the two plots. The visibility of the fringes of interference in the Wigner function is an indication of quantum correlations. }
\label{fig:wigner105}
\end{figure}

Theoretically, the Wigner function for the OAM state of a photon has been defined in the discrete cylinder $\mathbb{Z} \times C_{1}$ ($C_{1}$ is the unitary circle) representing the phase space for the OAM operator and its canonically conjugate operator $\hat{\theta}$.
%This unusual phase space arise because of the OAM quantization for the photons. 
In Ref.~\cite{wig} it has been shown that the study of the Wigner function for an OAM state gives information both on the various OAM eigenstates involved in the description of the state and on their relative phase. We define the two-mode Wigner function as
\begin{equation}
W(l_{A},\theta_{A},l_{B},\theta_{B})= \text{Tr}[\hat{w}(l_{A},\theta_{A})\otimes\hat{w}(l_{B},\theta_{B})\rho_{AB}],
\end{equation}
where we have introduced the kernel $\hat{w}(l_{I},\theta_{I})~(I=A,B)$ mapping quantum states in phase space~\cite{wig} in a way completely analogous to the continuous position-momentum phase space. In Fig.~\ref{fig:wigner105} {\bf a)} we compare the Wigner function $W(0,\theta,0,0)$ associated with $\rho_{AB}(t)$ when the entanglement transfer has not occurred [wide blue circle] to what is achieved at long-enough $t$, where the map $\hat{\cal M}_t$ has been implemented [inner (red) butterfly-like structure]. The difference due to the coherence established between two-mode orthogonal eigenstates  of the OAM operator in the entangled state $\rho_{AB}$ is striking.  Moreover, from Fig.~\ref{fig:wigner105} {\bf b)} and {\bf c)}, we find that $W(0,\theta_{A},0,\theta_{B})$ exhibits oscillations whose amplitude depends on the relative phase between the states $\ket{B}$ and $\ket{D}$. The curves shown in panels {\bf b)} and {\bf c)} are associated with the same interaction time as in the circular and burtterfly-like structures shown in panel {\bf a)}, respectively, which demonstrates that the oscillation amplitudes depend on the populations of these states. Remarkably, the analysis of $W(l_{A},\theta_{A},l_{B},\theta_{B})$ for incoherent superpositions of OAM eigenstates results in a flat distribution. 
%The reason is that in the single BEC there is not superposition of different OAM eigenstates but only an incoherent mixture as one can easly see from the single BEC reduced density matrix.
Therefore, by experimentally reconstructing the Wigner function, one can determine the inference of entanglement in the associated OAM state. Any deviation from flat distributions typical of incoherent superpositions implies coherence in the bipartite state, although not entanglement. Methods based on the inverse Radon transform~\cite{OHT} could then be used in order to achieve full information on the state and, eventually, the entanglement set by the transfer mechanism. 
%and comparing it with te expected behavior for to the  it is thus possible to know if the two BECs are entangled by comparing their Wigner function with the one of an inchoerent states which turn out to be a simple flat distribution.

%In this Section we highlight methods for the inference of the vortex entanglement described above. Our approach is based on the reverse of the process addressed here for light-to-BEC entanglement transfer. Suppose that an experiment has been run where two BECs have been entangled by means of the very same mechanism described here. The pump fields have thus been turned off (or set far off-resonant with respect to the frequency of the transitions they guide) so as not to perturb the entangled states of the vortices. The time-reversal nature of our protocol makes intuitive to understand that, if we now turn such pumps on again, photons will be scattered into two Laguerre-Gauss modes at the frequency of the $\ket{0}_{I}-leftrightarrow\ket{e,e'}$ transition, thus {\it writing back} the vortex state onto light fields. 

\section{Conclusions and outlook}
\label{conclusioni}
We have shown that vortex states of spatially remote, non-isotropically trapped BECs can be entangled by means of bilocal OAM-transfer processes and quantum correlated photonic resources. The amount of vortex entanglement set by our scheme can be quite considerable and appears to be limited only by the zero-OAM carrying component in the photonic resource. While such a bottleneck can be actively bypassed by means of post-selection, as described in Sec.~\ref{Transfer}, we are currently working on a modification of our protocol based on the use of other forms of the entangled photon-pair~\cite{leuchs}. The difference between the two-vortex state achieved by our scheme and a classical admixture of OAM eigenstates (without coherence and, thus, entanglement) can be revealed by a straigthforward {\it state-retrieval} process and the reconstruction of the OAM-state Wigner function. We believe that the superfluid phase of a BEC, together with virtually frictionless rotational states of light-induced vortices,  can be reliably exploited in order to set a promising scenario for the storage of quantum information and the distribution of quantum correlated channels for communication. 

\acknowledgments
We acknowledge support from Science Foundation Ireland (grant number 05/IN/I852). NLG is supported by the IRCSET through the Embark initiative No. RS/2000/137. MP thanks EPSRC (E/G004579/1) for financial support.

\renewcommand{\theequation}{A-\arabic{equation}}
\setcounter{equation}{0}
\section*{APPENDIX A}
\label{espliciti}
In this Appendix we present the calculations which lead to the expression of the effective Hamiltonian in Sec.~\ref{modello}. The first step is the adiabatic elimination of the excite triplet $\{\ket{E}$,$\ket{e}$,$\ket{e^{'}}\}$, which we perform by defining the rotating-picture new matter field operators $\hat{\psi}_{I,e}= \hat{\tilde \psi}_{I,e} e^{-i \omega_{l}t}$, $\hat{\psi}_{I,e^{'}}=\hat{\tilde \psi}_{I,e^{'}} e^{-i \omega_{-l}t}$ e $\hat{\psi}_{I,E}= \hat{\tilde \psi}_{I,E} e^{-i \omega_{0}t}$ with $\{\hat{\tilde \psi}_{I,E},\hat{\tilde \psi}_{I,e},\hat{\tilde \psi}_{I,e^{'}}\}$ taken as time-independent. Such operators are then used in the Heisenberg equations for the dynamics of the excited-triplet states. One finds
\begin{equation}
\label{eq:adiab}
\begin{aligned}
\hat{\tilde \psi}_{I,E}(\mathbf{r}) &=\frac{\chi_{I,0}\mathscr{A}_{n_{I},0}(\mathbf{r})}{\hbar\Delta_{I,0}}\;e^{i\omega_{0}t}\,\hat{c}_{n_{I},0}\,\hat{\psi}_{I,0}(\mathbf{r}),\\
\hat{\tilde \psi}_{I,e}(\mathbf{r}) &= \frac{\Omega_{I}}{\Delta_{I}} \;e^{i\mathbf{k}_{I}\cdot\mathbf{r}}e^{-i (\omega_{I}-\omega_{l})t}\,\hat{\psi}_{I,l}(\mathbf{r})\\
&+\frac{\chi_{I,l}\mathscr{A}_{n_{I},l}(\mathbf{r})}{\hbar\Delta_{I}}\;e^{i\omega_{l}t}\;\hat{c}_{n_{I},l}\,\hat{\psi}_{I,0}(\mathbf{r}),\\
\hat{\tilde \psi}_{I,e^{'}}(\mathbf{r}) &= \frac{\Omega_{I}}{\Delta_{I}} \;e^{i\mathbf{k}_{I}\cdot\mathbf{r}}\;e^{-i (\omega_{I}-\omega_{-l})t}\,\hat{\psi}_{I,-l}(\mathbf{r})\\
&+\frac{\chi_{I,-l}\mathscr{A}_{n_{I},-l}(\mathbf{r})}{\hbar\Delta_{I}}\;e^{i\omega_{-l}t}\,\hat{c}_{n_{I},-l}\,\hat{\psi}_{I,0}(\mathbf{r}),
\end{aligned}
\end{equation}
where we have defined the single photon detunings $\Delta_{I,0}=\omega_{0}-\omega_{I,E}$, $\Delta_{I}=\omega_{l}-\omega_{I,e}$. In order to explicitly include the chirped two-photon Raman detunings, which are crucial in the stabilization of the transfer process, we define new field operators in a rotating frame defined by the Hermitian operator $\hat{O}\!=\!\sum_{I,k}\omega_{I}\left(\hat{\psi}_{I,k}^{\dagger}(\mathbf{r}, t)\hat{\psi}_{I,k}(\mathbf{r}, t)-\hat{c}_{n_{I},k}^{\dagger}\hat{c}_{n_{I},k}\right)$. Explicitly
%\begin{equation}
%\label{eq:transf}
%\begin{aligned}
$\hat{\tilde \psi}_{I,k}(\mathbf{r}, t)=e^{i\hat{O}t}\hat{\psi}_{I,k}(\mathbf{r},t)e^{-i\hat{O}t}=e^{-i \omega_{I}t}\hat{\psi}_{I,k}(\mathbf{r}, t)$, while the photonic operators become $\hat{\tilde c}_{n_{I},k}=e^{i\hat{O}t}\hat{c}_{n_{I},k}e^{-i\hat{O}t}=e^{i \omega_{I}t}\hat{c}_{n_{I},k}$.
%,\,\hat{\tilde c}_{I,m}(t)=e^{i\hat{O}t}\hat{c}_{I,m}(t)e^{-i\hat{O}t}e^{i \omega_{I}t}=\hat{c}_{I,m}$
%\end{aligned}
%\end{equation}
Defining the free Hamiltonians $H_0^{I,k}=-\hbar^{2}\nabla_{I}^{2}/(2m)+V_{I,k}(\mathbf{r})$ and with the help of Eqs. ~(\ref{eq:adiab}) we finally get
\begin{equation}
\begin{aligned}
&i\hbar \frac{d}{dt}\hat{\tilde \psi}_{I,0}(\mathbf{r}) =[\hat{H}_{0}^{I,0} +\hbar\omega_{I}+\sum_{j}\eta_{j,0}^{I} \hat{\tilde \psi}_{I,j}^{\dagger}(\mathbf{r}) \hat{\tilde \psi}_{I,j}(\mathbf{r})\\
&+\frac{\chi_{I,0}^{2}\modq{\mathscr{A}_{n_{I},0}(\mathbf{r})}}{\hbar\Delta_{I,0}}\hat{\tilde c}_{n_{I},0}^{\dagger}\hat{\tilde c}_{n_{I},0}+\frac{\chi_{I,l}^{2}\modq{\mathscr{A}_{n_{I},l}(\mathbf{r})}}{\hbar\Delta_{I}}\hat{\tilde c}_{n_{I},l}^{\dagger}\hat{\tilde c}_{n_{I},l}\\
&+\frac{\chi_{I,-l}^{2}\modq{\mathscr{A}_{n_{I},-l}(\mathbf{r})}}{\hbar\Delta_{I}}\hat{\tilde c}_{n_{I},-l}^{\dagger}\hat{\tilde c}_{n_{I},-l}]\hat{\tilde\psi}_{I,0}(\mathbf{r})\\
&+\frac{\Omega_{I,l}\chi_{I,0}^{*}}{\Delta_{I}}\mathscr{A}_{n_{I},l}^{*}(\mathbf{r}) e^{i\mathbf{k}_{I} \cdot \mathbf{r}} \hat{\tilde c}_{n_{I},l}^{\dagger}\hat{\tilde \psi}_{I,l}(\mathbf{r})\\
&+\frac{\Omega_{I}\chi_{I,-l}^{*}}{\Delta_{I}}\mathscr{A}_{n_{I},-l}^{*}(\mathbf{r}) e^{i\mathbf{k}_{I} \cdot \mathbf{r}}\hat{\tilde c}_{n_{I},-l}^{\dagger} \hat{\tilde \psi}_{I,-l}(\mathbf{r}),
\end{aligned}
\end{equation}
\begin{equation}
\begin{aligned}
&i\hbar \frac{d}{dt}\hat{\tilde \psi}_{I,l}(\mathbf{r}) =[\hat{H}_{0}^{I,l}+\hbar(\omega_{l}-\delta_{l}-\tilde \omega_{l}) +  \frac{\hbar \modq{\Omega_{I}}}{\Delta_{I}}\\
&+\sum_{j}\eta_{j,l}^{I} \hat{\tilde \psi}_{I,j}^{\dagger}(\mathbf{r}) \hat{\tilde \psi}_{I,j}(\mathbf{r})]\hat{\tilde \psi}_{I,l}(\mathbf{r})\\
&+\frac{\Omega_{I}^{*}\chi_{I,l}}{\Delta_{I}} \mathscr{A}_{n_{I},l}(\mathbf{r})e^{-i\mathbf{k}_{I} \cdot \mathbf{r}} \hat{\tilde c}_{n_{I},l}\hat{\tilde \psi}_{I,0}(\mathbf{r}),
\end{aligned}
\end{equation}
\begin{equation}
\begin{aligned}
&i\hbar \frac{d}{dt}\hat{\tilde c}_{n_{I},0}\!=\![-\hbar \omega_{I}\!+\!\int_{{\cal V}_{I}}\!d\mathbf{r}\frac{\chi_{I,0}^{2}\modq{\mathscr{A}_{n_{I},0}(\mathbf{r})}}{\hbar\Delta_{I,0}}\hat{\psi}_{I,0}^{\dagger}(\mathbf{r}) \hat{\psi}_{I,0}(\mathbf{r})]\hat{\tilde c}_{n_{I},0}
\end{aligned}
\end{equation}
\begin{equation}
\begin{aligned}
&i\hbar\frac{d}{dt}\hat{\tilde c}_{n_{I},l}\!=\![-\hbar \omega_{I}\!+\!\int_{{\cal V}_{I}}\!d\mathbf{r}\frac{\chi_{I,l}^{2}\modq{\mathscr{A}_{n_{I},l}(\mathbf{r})}}{\hbar\Delta_{I}} \hat{\psi}_{I,0}^{\dagger}(\mathbf{r}) \hat{\psi}_{I,0}(\mathbf{r})]\hat{\tilde c}_{n_{I},l}\\
&\!+\!\int_{{\cal V}_{I}}\!d \mathbf{r} \quad\frac{\Omega_{I} \chi_{I,l}^{*}}{\Delta_{I}} \mathscr{A}_{n_{I},l}^{*}(\mathbf{r})e^{i\mathbf{k}_{I} \cdot \mathbf{r}} \hat{\tilde \psi}_{I,0}^{\dagger}(\mathbf{r}) \hat{\tilde \psi}_{I,l}(\mathbf{r}).
\end{aligned}
\end{equation}
From these expressions it is straightforward to define the effective interaction Hamiltonian $\hat{H}_{{\it eff}}$ used throughout the paper.


\begin{thebibliography}{99}
\bibitem{notaReferee} We have verified the validity of the description given by the adiabatic Hamiltonian $\hat{H}_{eff}$ by numerically solving the full problem represented by Eq.~(\ref{eq:fullham}) for the same initial state and using parameters validating the use of the effective coupling model. We have checked that the normalization of the state vector spanned by the states in Eq.~(\ref{chiusi}) is mantained within the time window used for our study, thus showing negligible leakage outside the relevant sector of the Hilbert space.
\bibitem{commentoT} By means of continuously pumped cavity-enhanced SPDC, which is a customary method for the generation of photonic resources for atomic memories and quantum repeaters, one can have highly-monochromatic twin-beam states in spatially very spread single-photon wavepackets carrying the desired value of angular momentum. Such states can be treated, for all practical purposes, as plane waves in our calculations. However, it is straightforward to adapt our formalism to the case of  pulsed SPDC by making the Rabi frequencies appearing in $\hat{H}_{eff}$ explicitly time dependent, so as to incorporate the form of the photonic wavepacket.
\end{thebibliography}
\end{document}